\documentclass[12pt,preprint]{elsart}
\usepackage{graphicx}
\usepackage{epsfig}
\usepackage{amsmath}
\usepackage{amssymb}

\def\NPB{{\em Nucl. Phys.} B}

\def\PR#1#2#3{{\rm Phys. Rep.} {\bf{#1}} (#2) #3}
\def\PRD#1#2#3{{\rm Phys. Rev.} {\bf{D#1}} (#2) #3}

\def\NPB#1#2#3{{\rm Nucl. Phys.} {\bf{B#1}} (#2) #3}

\def\PRL#1#2#3{{\rm Phys.~Rev.~Lett.} {\bf{#1}} (#2) #3}

\def\APP#1#2#3{{\rm Astropart.~Phys.} {\bf{#1}} (#2) #3}
\def\APJ#1#2#3{{\rm Astrophys.~J.} {\bf{#1}} (#2) #3}

\def\AA#1#2#3{{\rm A\&A} {\bf{#1}} (#2) #3}
\def\MNRAS#1#2#3{{\rm MNRAS} {\bf{#1}} (#2) #3}
\def\NATURE#1#2#3{{\rm Nature} {\bf{#1}} (#2) #3}
\def\JCAP#1#2#3{{\rm JCAP} {\bf{#1}} (#2) #3}
\begin{document}
\runauthor{F.~A. Aharonian et al.}

\begin{frontmatter}

\title{Observations of the Sagittarius Dwarf galaxy by the H.E.S.S. experiment\\ and search for a Dark Matter signal}

{\small \author{ F. Aharonian$^{1}$},
 \author{ A.G.~Akhperjanian$^{2}$},
  \author{ A.R.~Bazer-Bachi$^{3}$},
 \author{ M.~Beilicke$^{4}$},
 \author{ W.~Benbow$^{1}$},
 \author{ D.~Berge$^{1,}$$^{a}$},
 \author{ K.~Bernl\"ohr$^{1,5}$},
 \author{ C.~Boisson$^{6}$},
 \author{ O.~Bolz$^{1}$},
 \author{ V.~Borrel$^{3}$},
% ...
}
\newpage
\begin{flushleft}
\footnotesize{{\small $^{\star}$ Corresponding author :
moulin@in2p3.fr; DAPNIA/DSM/CEA, CE Saclay, F-91191 Gif-sur-Yvette
Cedex, France
\address{$^{1}$
Max-Planck-Institut f\"ur Kernphysik, P.O. Box 103980, D 69029
Heidelberg, Germany}
\address{$^{2}$
 Yerevan Physics Institute, 2 Alikhanian Brothers St., 375036 Yerevan,
Armenia
}
\address{$^{3}$
Centre d'Etude Spatiale des Rayonnements, CNRS/UPS, 9 av. du Colonel Roche, BP
4346, F-31029 Toulouse Cedex 4, France
}
\address{$^{4}$
Universit\"at Hamburg, Institut f\"ur Experimentalphysik, Luruper Chaussee
149, D 22761 Hamburg, Germany
}
\address{$^{5}$
Institut f\"ur Physik, Humboldt-Universit\"at zu Berlin, Newtonstr. 15,
D 12489 Berlin, Germany
}
\address{$^{6}$
LUTH, UMR 8102 du CNRS, Observatoire de Paris, Section de Meudon, F-92195 Meudon Cedex,
France
}
% ...
\address{$^{a}$ now at CERN, Geneva, Switzerland}
% ...
}}
\end{flushleft}

\newpage

\begin{abstract}
Observations of the Sagittarius dwarf spheroidal (Sgr dSph) galaxy
were carried out with the H.E.S.S. array of four imaging air
Cherenkov telescopes in June 2006. A total of 11 hours of high
quality data are available after data selection. There is no
evidence for a very high energy $\gamma$-ray signal above the
energy threshold at the target position. A 95\% C.L. flux limit of
$\rm 3.6 \times 10^{-12} cm^{-2}s^{-1}$ above 250 GeV has been
derived.
 Constraints on the velocity-weighted cross section $\rm \langle \sigma v \rangle$ are
calculated in the framework of Dark Matter particle annihilation
using realistic models for the Dark Matter halo profile of
Sagittarius dwarf galaxy. Two different models have been
investigated encompassing a large class of halo types. A 95\% C.L.
exclusion limit on $\rm \langle \sigma v \rangle$ of the order of
$\rm 2 \times 10^{-25} cm^3s^{-1}$  is obtained for a core profile
in the 100 GeV - 1 TeV neutralino mass range.

\end{abstract}


\end{frontmatter}


\newpage
%
\section{Introduction}
\label{sec:intro} Astrophysical and cosmological observations
provide a substantial body of evidences for the existence of Cold
Dark Matter (CDM) although its nature remains still unknown. It is
commonly assumed that CDM is composed of yet undiscovered
non-baryonic particles for which plausible candidates are Weakly
Interacting Massive Particles (WIMPs). In most theories,
candidates for CDM are predicted in theories beyond the Standard
Model of particle physics~\cite{bertone}. The indirect detection
of Dark Matter (DM) annihilation may bring new insights to probe
the astrophysical nature of Dark Matter.  To a large extent
complementary to direct searches, the indirect detection enables
to search for DM outside the solar system. Indeed, it may give
detailed morphology features that may constrain the DM halo
profile. The annihilation of WIMPs into $\gamma$-rays may lead to
detectable very high energy (VHE, E $>$ 100 GeV) $\gamma$-ray
fluxes above background via continuum emission from hadronization
of gauge bosons and heavy quarks, or $\gamma$-ray lines through
loop-induced processes. The H.E.S.S. (High Energy Stereoscopic
System) array of Imaging Atmospheric Cherenkov Telescopes (IACTs),
designed for high sensitivity measurements in the 100 GeV - 10 TeV
energy regime, is a suitable instrument to
detect VHE $\gamma$-rays and investigate their possible origin.\\
Various astrophysical systems ranging from local objects in the
galactic halo to galaxy cluster scales have been considered as
targets for DM annihilation $\gamma$-ray studies. Interacting with
baryonic matter only through gravity, the WIMPs are expected to
concentrate at the centre of high density region. A probable
candidate is the elliptical galaxy M87 at the centre of the Virgo
cluster~\cite{baltz}. However, the temporal variability of the
H.E.S.S. signal~\cite{science} excludes the bulk of the signal in
the TeV range to be of a dark matter origin. Prospects of indirect
detection from the Andromeda galaxy M31~\cite{fornengo} and the
Large Magellanic Cloud~\cite{tasitsiomi} have been also
investigated. H.E.S.S. observations in the direction of the
Galactic Center (GC) have revealed a source of VHE $\gamma$-ray
emission (HESS J1745-290). The measured spectrum is difficult to
reconcile with a DM interpretation and does not match the expected
annihilation spectrum. The very high energy cut-off above 7 TeV
requires an uncomfortably massive DM particle~\cite{prl}.
Moreover, standard astrophysical objects found in the region such
as the supermassive black hole Sgr A* or the recently discovered
plerion G359.95-0.04, can easily account for the observed signal.
The shell-type supernova remnant Sgr A East is unlikely to be
associated to the signal~\cite{vanEldik}. Besides HESS J1745-290,
the deep observations carried out with H.E.S.S. in the GC region
have highlighted the existence of a diffuse component along the
galactic ridge~\cite{nature}. This extended TeV emission may be
explained by cosmic ray interactions inside the central molecular
zone leading to diffuse astrophysical backgrounds that might hide
the exotic signal~\cite{aharonian}. Future indirect DM searches in
this region will have to overcome this challenging background. In
contrast, dwarf galaxies may be a less complex environment because
of the reduced amount of
gas in such small systems~\cite{evans}.\\
Dwarf Spheroidal galaxies (dSph) such as Sagittarius dwarf or
Canis Major, discovered recently in the Local Group, are among the
most extreme DM-dominated environments. Indeed, measurements of
roughly constant radial velocity dispersion of stars imply large
mass to luminosity ratios~\cite{wilkinson}. Nearby dwarfs are
ideal
astrophysical 
probes of the nature of DM as they usually consist of a stellar
population with no hot or warm gas, no cosmic ray population and
little dust. The Sagittarius dwarf Spheroidal galaxy (Sgr dSph) is
the next-to-last Galactic satellite galaxy
discovered~\cite{ibata0}.
 The core of Sgr dSph is located at l =
5.6$^{\circ}$ and b = -14$^{\circ}$ in galactic coordinates at a
distance of about 24 kpc from the Sun~\cite{majewski}. Sgr dSph
has made at least ten Milky Way crossings it should thus contain a
substantial amount of DM to avoid to have been entirely disrupted.
Latest velocity dispersion measurements on M giant stars with
2MASS yields a light to mass ratio of about 25~\cite{majewski2}.
The Sgr dSph core is positioned behind the bulge of Milky Way but
outside the Galactic plane, thus reduced foreground  $\gamma$-ray
contaminations are expected. The luminous density profile of Sgr
dSph has two components~\cite{monaco}. The compact component,
namely the core, is characterized by a size of about 3 pc FWHM,
which corresponds to a point-like region for H.E.S.S. This is the
DM annihilation region from which $\gamma$-ray signal may be
expected. A diffuse component is well fitted by a King model with
a characteristic size of 1.6 kpc.\\
In this paper, we present the observation of the Sagittarius dwarf
galaxy by the H.E.S.S. array of Imaging Atmospheric Cherenkov
Telescopes based on a dataset collected in June 2006. A careful
modeling of the Dark Matter halo using the latest measurements on
the structural parameters of Sagittarius dwarf is presented to
derive constraints on the WIMP velocity-weighted annihilation
rate.

\section{Search of VHE $\gamma$-rays from observations of Sagittarius Dwarf by H.E.S.S.}
\label{sec:analysis}
\subsection{The H.E.S.S. instrument}
H.E.S.S. (High Energy Stereoscopic System) is an array of four
imaging atmospheric Cherenkov telescopes located in the Khomas Highlands
of Namibia at an altitude of 1800 m above sea level. Each
telescope 
consists of an optical reflector of about 107 m$^2$ effective area
composed of 382 round mirrors arranged on a Davis-Cotton
mount~\cite{bernlohr}. The interaction of the primary $\gamma$-ray
in the Earth's upper atmosphere initiates an electromagnetic
shower. The reflector collects the Cherenkov light emitted by the
charged particles composing this shower, and focuses it onto a
camera comprising 960 fast photomultipliers (PMTs) of individual
field of view of 0.16$^{\circ}$ diameter~\cite{vincent}. Each tube
is equipped with Winston cones to limit the field of view of each
PMT and minimize the background light. The total field of view of
the H.E.S.S. instrument is 5$^{\circ}$ in diameter. The
stereoscopy technique used in the imaging atmospheric Cherenkov
telescopes allows for accurate reconstruction of the direction and
energy of the primary $\gamma$-rays as well as an efficient
rejection of the background induced by cosmic ray
interactions~\cite{trigger}. The energy threshold of H.E.S.S. at
zenith before selection cuts moved from 100 GeV at the
commissioning of the experiment in 2003 to 160 GeV due to the
degradation of the optical performance in 2006. The point source
sensitivity is better than $\rm 2\times 10^{-13} cm^{-2}s^{-1}$
above 1 TeV for a 5$\sigma$ detection in 25 hours~\cite{crabe}.

\subsection{Dataset}
The observations of the Sgr dSph were taken in June 2006 with
zenith angles ranging from 7$^{\circ}$ to 43$^{\circ}$ around an
average value of 19$^{\circ}$. Data were taken in 28-minute
observation runs in the wobble mode method with pointing
directions offset by an angular distance of typically
$\pm$0.7$^{\circ}$ from the nominal target position. The dataset
suitable for analysis was selected using the standard run
selection procedure~\cite{crabe}, which in particular removes the
data taken under bad atmospheric conditions. A total of 25 runs
out of 26 are selected for the analysis. After calibration of the
raw shower images from PMT signals~\cite{aharonian0}, two
independent reconstruction techniques were combined to select
$\gamma$-ray events and reconstruct their direction and energy.
The first one uses the Hillas moment method~\cite{aharonian1}. The
second analysis referred hereafter as ``Model Analysis'', is based
on the pixel-per-pixel comparison of the shower image with a
template generated by a semi-analytical shower development model.
The event reconstruction relies on a maximum likelihood method
which uses available pixels in the camera, without requirement for
an image cleaning~\cite{denauroi0,denauroi1}. The reconstructed
shower parameter (energy, impact, direction and primary
interaction point) are obtained as a product of the minimization
procedure.
 The separation between $\gamma$ candidates and hadrons is done
 using a combination of the Model goodness-of-fit
 parameter~\cite{denauroi1} and the Hillas mean scaled width and length
 parameters, which results in an improved background
 rejection~\cite{crabe}.
Standard cuts on the width and the length of Hillas ellipses
combined with the goodness-of-fit are used to suppress the
hadronic background~\cite{aharonian1}. An additional cut on the
primary interaction depth is used to improve background rejection.
Both methods yield a typical energy resolution of 15\% above
energy threshold. In the Model Analysis, the angular resolution at
the 68\% containment radius is found to be better than
0.06$^{\circ}$ per $\gamma$-ray.
\subsection{Data analysis}
\begin{figure}[p]
\begin{center}
\mbox{\hspace{0cm}\includegraphics[scale=0.5]{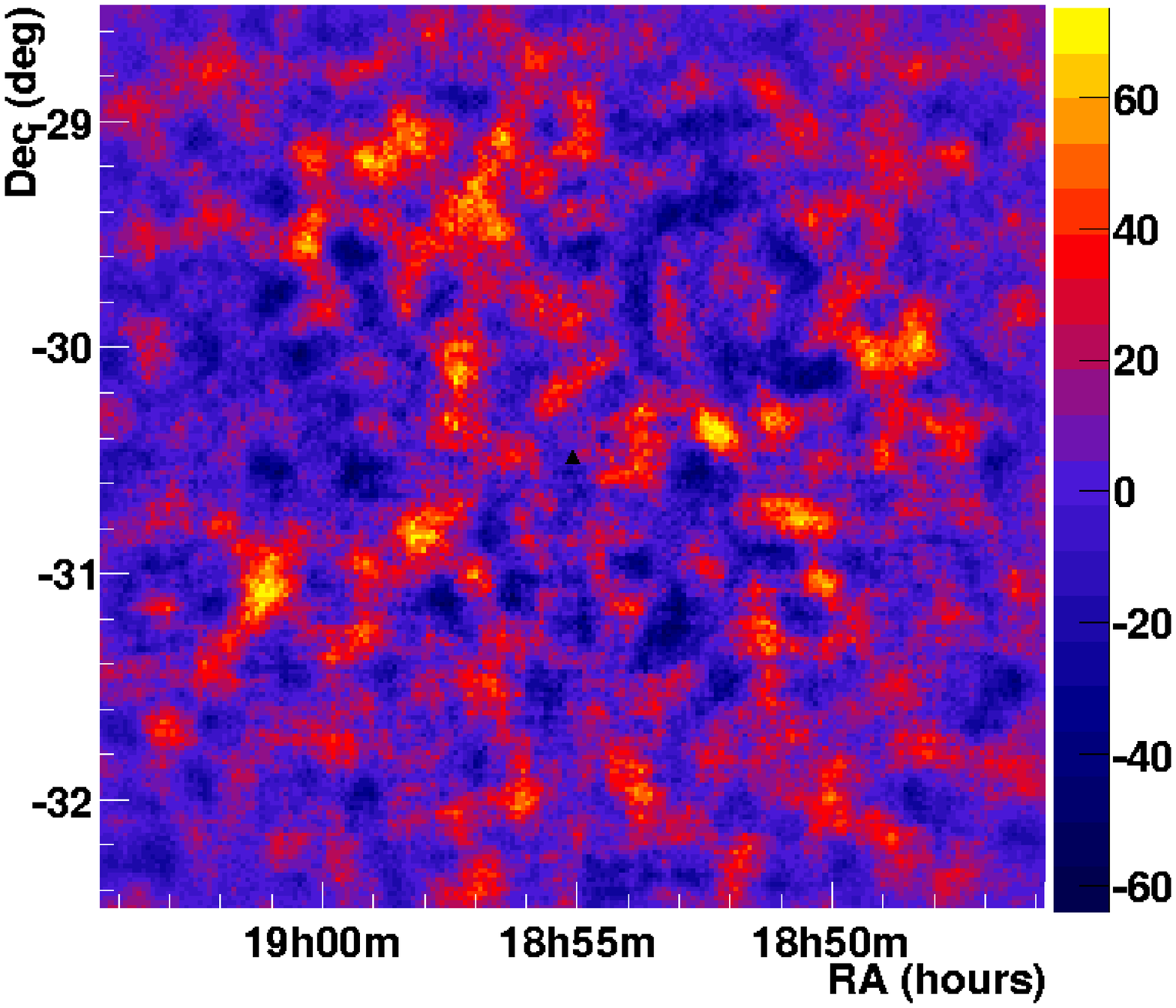}}
\caption{Sky map of the $\gamma$-ray candidates with an
oversampling radius of 0.14$^{\circ}$. No excess is observed at
the target position $\rm (RA = 18h54m40s,Dec = -30d27m05s)$ in
equatorial coordinates (J2000) marked with a black triangle. Other
spots in the field of view are not significant.}
 \label{fig:excessmap}
\end{center}
\end{figure}

\begin{figure}[p]
\begin{center}
\mbox{\hspace{0cm}\includegraphics[scale=0.5]{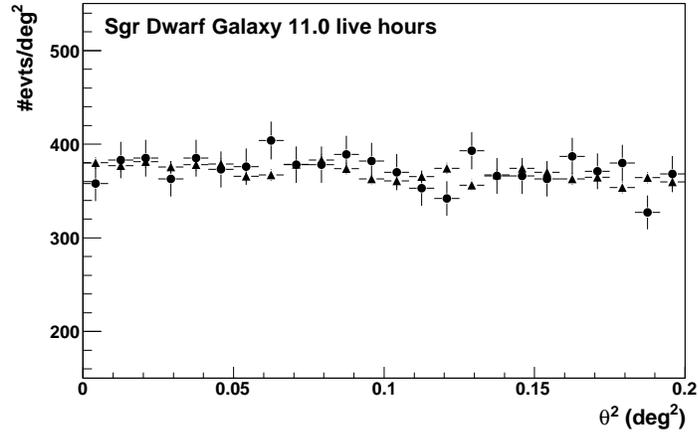}}
\caption{$\rm \theta^2$ radial distribution of the ON and OFF
events for $\gamma$-ray like events from the target position $\rm
(RA = 18h54m40s, Dec =-30d27m05s)$ (black dots). Estimated
background calculated as explained in the text is shown (black
triangles). No excess is seen at small $\theta^2$ value.}
 \label{fig:theta2}
\end{center}
\end{figure}

The on-source signal is defined by integrating all the events with
angular position $\theta$ in a circle around the target position
with a radius of $\theta_{cut}$. The target position is chosen
according to the photometric measurements of the Sgr dSph luminous
cusp showing that the position of the centre corresponds to the
centre of the globular cluster M54~\cite{monaco1}. The target
position is thus found to be $\rm (RA = 18^h55^m59.9s, Dec =
-30d28'59.9'')$ in equatorial coordinates (J2000.0) or $\rm (l =
5^{\circ}41'12.9'',b = -14^{\circ}16'29.8'')$ in Galactic
coordinates. The signal coming from Sgr dSph is expected to come
from a region of 1.5 pc, about 30'', much smaller the H.E.S.S.
point spread function (PSF). A $\theta_{cut}$ value of
0.14$^{\circ}$ suitable for a point-like source was therefore used
in the analysis. In case of a Navarro-Frenk-White (NFW) density
profile~\cite{nfw} for which $\rho$ follows r$^{-1}$ or a cored
profile~\cite{evans} folded with the point spread function (PSF)
of H.E.S.S., the integration region allows to retrieve a
significant fraction of the expected signal. See
Table~\ref{tab:table}. A cut on the image size of 60
photoelectrons is used to obtain a good sensitivity for weak
sources. In order to reduce systematic effects which affect images
close to the edges of the camera, only events reconstructed within
a maximum distance of 2.5$^{\circ}$ from the camera centre are
used for this analysis. The excess sky map is obtained by the
subtraction of a background model on the $\gamma$-ray candidate
sky distribution. The background level is estimated using the
ring-background method~\cite{puhlhofer} where the background rate
is calculated from the integration of $\gamma$-like events falling
in an annulus around the centre of the camera with identical
observation conditions and acceptances than that used for the
on-source region, which allows
an estimate of the background on every sky position.\\
The excess sky map of the $\gamma$-ray candidates is presented in
Fig.~\ref{fig:excessmap} in the RA/Dec J2000 coordinates centered
on the Sgr target position. For each bin, $\gamma$-ray like events
are summed within a radius of 0.14$^{\circ}$. No $\gamma$-ray
excess is found at the target position. Further, no significant
excess is observed anywhere else in the sky map. The $\theta^2$
distribution of the observed $\gamma$-ray events relative to the
target position, is presented in Fig.~\ref{fig:theta2} as well as
the background distribution. In order to check the robustness of
the results presented here, the dataset has been analyzed using
different analysis methods as the so-called Hillas and
model3D~\cite{lemoine} methods. No significant $\gamma$-ray excess
is detected
in the corresponding sky maps.\\
Since Fig.~\ref{fig:theta2} shows no $\gamma$-ray excess, we
derived the 95\% confidence level upper limit on the observed
number of $\gamma$-rays : $\rm N_{\gamma}^{95\%\, C.L.}$. The
limit is computed knowing the numbers of events above the energy
of 250 GeV in the signal region $\rm N_{ON} = 437$, in the
background region $\rm N_{OFF} = 4270$, and the ratio of the
off-source time to the on-source time $\rm\alpha = 10.1$. We use
the Feldman \& Cousins method~\cite{feldman} and obtain :
\begin{equation}
N_{\gamma}^{95\%\,C.L.} = 56\,.
\end{equation}
Given the acceptance of the detector for the observations of the
dSph Sgr, an 95\% confidence level upper limit on the $\gamma$-ray
flux is also derived :
\begin{center}
$\rm \Phi_{\gamma}(E_{\gamma} > 250\,GeV) < 3.6 \times 10^{-12}
\,cm^{-2}s^{-1}\,(95\%\,C.L.)$
\end{center}

\section{Predictions of $\gamma$-rays from Dark Matter annihilation}
\label{sec:predictions}
\subsection{Theoretical framework}
The annihilation of DM particles can generate $\gamma$-ray fluxes
through different processes depending on the particle physics
scenarios. Generally, WIMP annihilations will produce a continuum
of $\gamma$-rays with energies up to the WIMP mass issued from the
hadronization and decay of the cascading annihilation products,
predominantly from $\pi^0$'s generated in the quark jets. In the
R-parity conserving supersymmetric extensions of the Standard
Model, the lightest supersymmetric particle (LSP) is a stable
particle and is a good CDM candidate. The LSP is, in various SUSY
breaking models, the lightest neutralino $\tilde{\chi}$. Being
electrically neutral and colorless, it
 is among the best motivated candidates to account
for CDM~\cite{jungman}. In Minimal Supersymmetric Standard Model
(MSSM) scenarios~\cite{jungman}, the annihilation of neutralinos
can, on top of the hadronization continuum, produce also
monoenergetic spectral lines of $\gamma$-rays resulting from
loop-induced annihilation processes such as
$\tilde{\chi}\tilde{\chi}\rightarrow\gamma\gamma$,
$\tilde{\chi}\tilde{\chi}\rightarrow\gamma Z$,
$\tilde{\chi}\tilde{\chi}\rightarrow\gamma h$, even though these
are very challenging to detect experimentally due to the high
suppression of such final states. Beyond the Standard Model,
plausible candidates are provided by the universal extra dimension
(UED) theories. In Kaluza-Klein (KK) scenarios with KK-parity
conservation, the lightest Kaluza-Klein particle (LKP) is
stable~\cite{servant}, the best-motivated being the first KK mode
of the hypercharge gauge boson, $\tilde{B}^{(1)}$. In this case,
$\tilde{B}^{(1)}$ pairs annihilate preferentially into charged
lepton pairs which radiatively produce $\gamma$ with harder
spectra. Cascading decays of $q\bar{q}$ final states lead to
secondary
$\gamma$-rays~\cite{bergstrom2}.\\
The $\gamma$-ray flux from  annihilations of DM particles of mass
$m_{DM}$ accumulating in a spherical DM halo can be expressed in
the form :
\begin{equation}
\label{eqnp}
\frac{d\Phi(\Delta\Omega,E_{\gamma})}{dE_{\gamma}}\,=\frac{1}{4\pi}\,\underbrace{\frac{\langle
\sigma
v\rangle}{m^2_{DM}}\,\frac{dN_{\gamma}}{dE_{\gamma}}}_{Particle\,
Physics}\,\times\,\underbrace{\bar{J}(\Delta\Omega)\Delta\Omega}_{Astrophysics}
\end{equation}
as a product of a particle physics component with an astrophysics
component. The particle physics part contains $\langle \sigma
v\rangle$, the velocity-weighted annihilation cross section, and
$dN_{\gamma}/dE_{\gamma}$,  the differential $\gamma$-ray spectrum
summed over the whole final states with their corresponding
branching ratios. The astrophysical part corresponds to the
line-of-sight-integrated squared density of the DM distribution J,
averaged over the instrument solid angle integration region for
H.E.S.S. ($\Delta\Omega = 2\times 10^{-5}$ sr) :
\begin{equation}
\label{eqnj} J\,=\,\int_{l.o.s}\rho^2(r[s])ds \hspace{2cm}
\bar{J}(\Delta\Omega)\,=\,\frac{1}{\Delta\Omega}\,\int_{\Delta\Omega}
\,{\rm PSF}*J\, d\Omega
\end{equation}
where PSF is the point spread function of H.E.S.S.
\subsection{Modeling the Sagittarius dwarf Dark Matter halo}
The mass distribution of the DM halo of Sgr dwarf has been
described by plausible models taking into account the best
available measurements of the Sgr dwarf galactic structure
parameters. We have used two widely different models. The first
has a NFW cusped profile~\cite{nfw} with the mass density given by
:
\begin{equation}
\rho_{NFW}(r)\,=\,\frac{A}{r(r+r_s)^{2}}
\end{equation}
with A the normalization factor and $r_{s}$ the scale radius taken
from~\cite{evans}. Using Eq.~\ref{eqnj}, the value of $\bar{J}$
obtained with this model is reported in
Table~\ref{tab:table}. \\
We have also studied a core-type halo model  as in~\cite{evans}
characterized by the mass density :
\begin{equation}
\rho_{core}(r)\,=\,\frac{v_a^2}{4\pi G}\frac{3 r_c^2
+r^2}{(r_c^2+r^2)^2}
\end{equation}
where $r_c$ is the core radius and $v_a$ a velocity scale.
 However, we have tried to update the $v_{a}$ and $r_{c}$
values which were used in~\cite{evans}. By inserting in the Jeans
equation the luminosity profile of the Sgr dwarf core of the form
:
\begin{equation}
\nu (r)= \frac{\nu_{0}{r_c}^{2\alpha}}{(r_c^2+r^2)^{\alpha}}
\end{equation}
we estimated from the data of reference \cite{monaco1} $\alpha =
2.69\pm 0.10$ and $ r_c = 1.5\ \mbox{pc}.$ Note that the value of
$r_{c}$ is only an upper limit. The value of the central velocity
dispersion of Sgr Dwarf is
 $\sigma = 8.2\,\pm\, 0.3\,\rm km
s^{-1}$~\cite{zaggia}. We have assumed that the velocity
dispersion is independent of position. The value of $v_{a}$ is
then given by $v_{a} = \sqrt{\alpha}\,\sigma = 13.4\,\rm km
s^{-1}$. The cored model gives a very large value of $\bar{J},$
which is reported in Table~\ref{tab:table}. The third column of
Table~\ref{tab:table} gives the amount of signal expected in the
solid angle integration region $\rm \Delta\Omega = 2 \times
10^{-5}$ sr.\\
The value of $\rm \bar{J}$ in the cored model depends on various
parameters such as the radial dispersion velocity, the baryon
fraction in the core, the core radius and the velocity tensor
anisotropy. The central value of the radial dispersion velocity
was taken as 8.2 $\rm km s^{-1}$. Previous measurements report
11.4 $\rm km s^{-1}$~\cite{ibata} which would lead to an increase
of a factor 4 in $\rm \bar{J}$. Deviations from the asymptotically
flat rotation curve have been studied. The effect of the $\alpha$
parameter from~\cite{evans} leads to a 50\% increase or decrease
in $\rm \bar{J}$, according to the sign of $\alpha$. The
anisotropy in the velocity dispersion may lead to a reduction of a
factor 2 in the dark matter density in the central region with
respect to the case without anisotropy~\cite{merritt}. The baryon
mass fraction in the very central region can not be neglected.
However, the effect on $\rm \bar{J}$ turns out to be at most 50\%
due to the relatively fast increase of the mass-to-luminosity
ratio with radius. Finally, in our model, the ratio of the
luminous core radius to the dark matter core radius is $\rho = 1$.
In general, as emphasized in~\cite{pryor}, $\rho$ could be lower
than 1. $\rm \bar{J}$ is strongly dependent on the value of
$\rho$. Extending the dark halo to 200 pc would lower $\rm
\bar{J}$ by 2 orders of magnitude. Note also that our value of the
luminous radius is only an upper limit. Decreasing the luminous
radius leads to an increase of $\rm \bar{J}$.

\subsection{Sensitivity}
\begin{figure}[p]
\begin{center}
\mbox{\hspace{0cm}\includegraphics[scale=0.7]{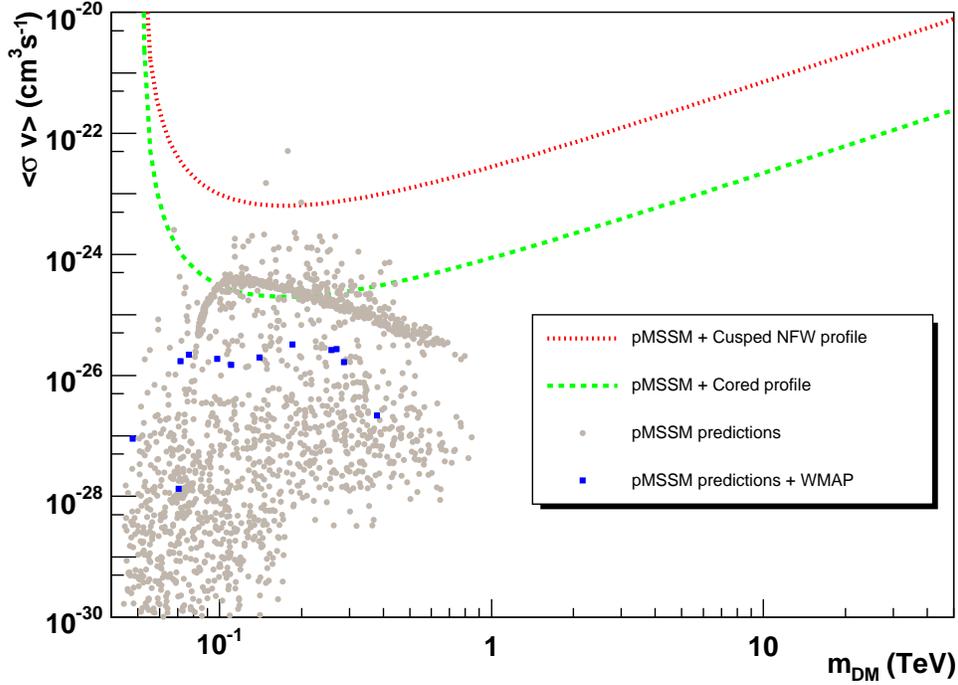}}
\caption{Upper limits at 95\% C.L. on $\langle \sigma v \rangle$
versus the DM particle mass in the case of a cusped NFW 
(red dotted line) and a cored (green dashed line) DM halo profiles
respectively. The pMSSM parameter space was explored with DarkSUSY
4.1~\cite{darksusy}, each point on the plot corresponding to a
specific model (grey point). Amongst these models, those
satisfying in addition the WMAP constraints on the CDM relic
density are overlaid as blue square (see text for details). The
limits in case of neutralino dark matter from pMSSM are derived
using the parametrisation from reference~\cite{bergstrom} for a
higgsino type neutralino annihilation $\gamma$ profiles.}
 \label{fig:sigmav}
\end{center}
\end{figure}

\begin{figure}[p]
\begin{center}
\mbox{\hspace{0cm}\includegraphics[scale=0.7]{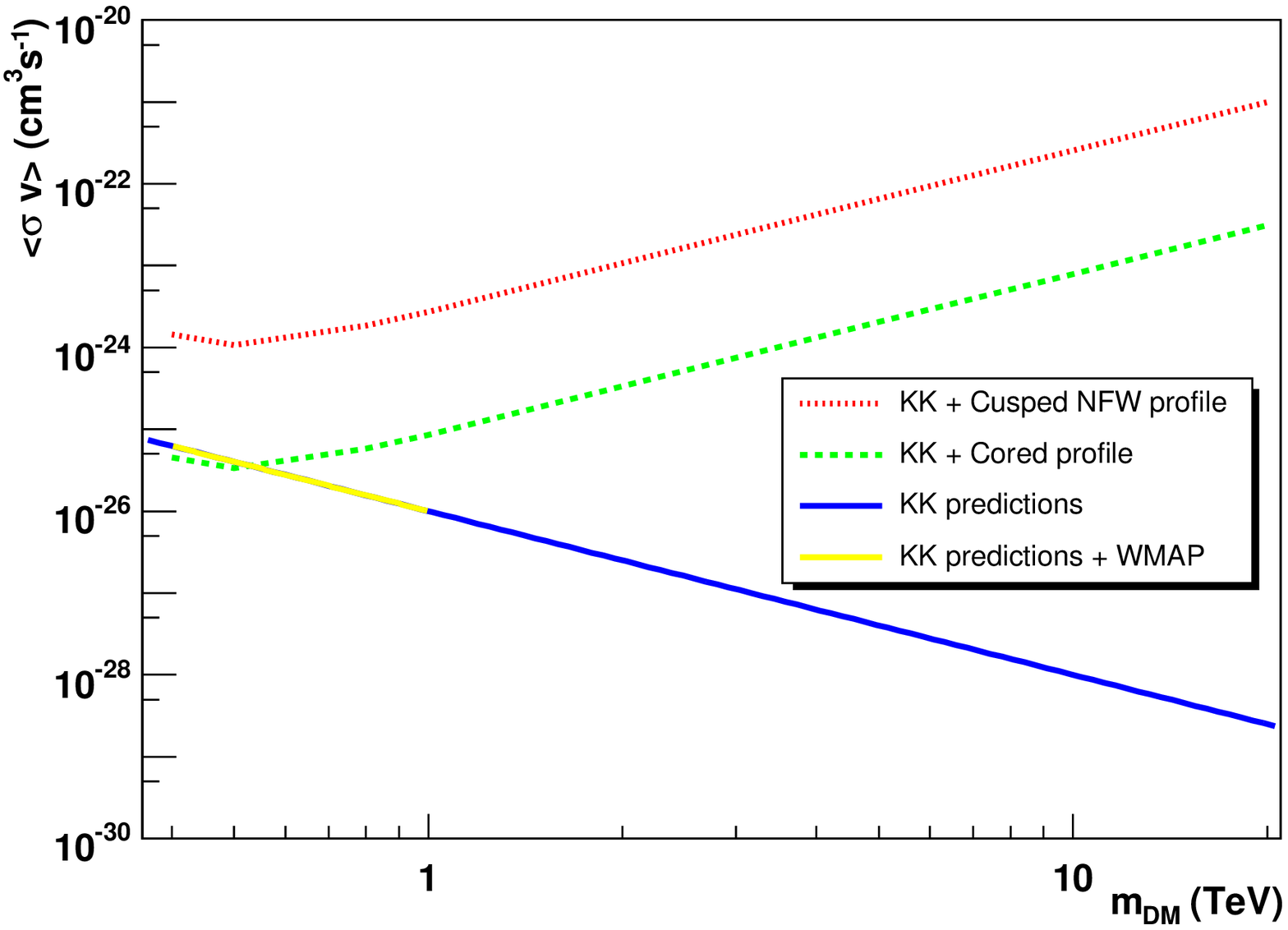}}
\caption{Upper limits at 95\% C.L. on $\langle \sigma v \rangle$
versus the DM particle mass in the  B$^{(1)}$ Kaluza-Klein
scenarios for a cusped NFW (red dotted line) and a cored (green
dashed line) DM halo profiles respectively. The blue line
corresponds to Kaluza-Klein models~\cite{servant}. Overlaid
(yellow line) are the KK models satisfying WMAP constraints on the
CDM relic density (see text for details). }
 \label{fig:sigmav2}
\end{center}
\end{figure}

For a given DM halo, the relevant quantities for DM particle
annihilation searches are the DM particle mass $m_{DM}$ and the
velocity-weighted cross section $\langle \sigma v \rangle$. With
the limit on the number of $\gamma$, N$_{\gamma}$, derived in
section 2.4, we can compute the limit on $\langle \sigma v
\rangle$ from H.E.S.S. results with the Sgr dwarf DM halo profile
modeled in section 3.2. N$_{\gamma}$ may be computed using the
formula :
\begin{equation}
\label{eqnng}
N_{\gamma}\,=\,T_{\mathrm{obs}}\,\int_{0}^{m_{DM}}A_{\mathrm{eff}}(E_{\gamma})\frac{d\Phi(\Delta\Omega,E_{\gamma})}{dE_{\gamma}}dE_{\gamma}
\end{equation}
where $\rm A_{eff}$ corresponds to the effective area of the
instrument obtained from Monte Carlo simulations as a function of
the zenith angle, the offset of the source from the pointing
direction, the energy of the event and the selection cuts. Using
the expression of the differential flux given in Eq.~\ref{eqnp}
with Eq.~\ref{eqnng} yields a 95\% C.L. exclusion limit on the
velocity-weighted cross section versus the DM particle mass for a
given halo profile as defined by :
\begin{equation}
\langle \sigma v \rangle_{min}^{95\%\,
C.L.}\,=\,\frac{4\pi}{T_{\mathrm{obs}}}\frac{m^2_{DM}}{\bar{J}(\Delta
\Omega)\Delta\Omega}\,\frac{N_{\gamma}^{95\%\,C.L.}}{\displaystyle
\int_{0}^{m_{DM}}A_{\mathrm{eff}}(E_{\gamma})\frac{dN_{\gamma}}{dE_{\gamma}}dE_{\gamma}}
\end{equation}
where $dN_{\gamma}/dE_{\gamma}$ is computed with a parametrization
of the differential continuum photon spectrum
from~\cite{bergstrom} for a higgsino-type neutralino.
Fig.~\ref{fig:sigmav} shows the limits in the case of a cored
(green dashed line) and cusped NFW (red dotted line) profile using
the value of $\bar{J}$ computed in section 3.2. Predictions for
phenomenological MSSM (pMSSM) models are displayed (grey points)
as well as those satisfying in addition the WMAP constraints on
the CDM relic density $\rm \Omega_{CDM} h^2$ denoted as blue
points. The values allowed by WMAP are taken to lie in the range
$\rm 0.09 \le \Omega_{CDM} h^2 \le 0.11$. The SUSY models are
calculated with DarkSUSY4.1~\cite{darksusy} in pMSSM framework and
characterized by a basic set of independent parameters  : the
higgsino mass parameter $\mu$, the gaugino mass parameter M$_2$,
the CP-odd Higgs mass M$_A$, the common scalar mass m$_0$ , the
trilinear couplings A$_{t,b}$ and the Higgs vacuum expectation
value ratio $\tan \beta$. The set of parameters for a given model
is randomly chosen in a parameter region encompassing a large
class of pMSSM models, as described in
Tab.~\ref{tab:table2}.\\
In the case of a cusped NFW profile, the H.E.S.S. observations do
not establish severe constraints on the velocity-weighted cross
section. For a cored profile, due to a higher central density,
stronger constraints are derived and some pMSSM models can be excluded in the upper part of the pMSSM scanned region. \\
In the case of KK dark matter, the differential $\gamma$ spectrum
is parametrized using Pythia~\cite{pythia} simulations and
branching ratios from~\cite{servant}. Predictions for the
velocity-weighted cross section of B$^{(1)}$ dark matter particle
are performed using the formula given in~\cite{sigmavKK}. In this
case, the expression for $\langle \sigma v \rangle$ depends
analytically on the B$^{(1)}$ mass square. Fig.~\ref{fig:sigmav2}
shows the sensitivity of H.E.S.S. in the case of Kaluza-Klein
models where the hypercharge boson B$^{(1)}$ is the LKP, for a
cored (green solid line) and a cusped NFW (red solid line) profile
respectively using the value of $\rm \bar{J}$ computed in section
3.2. With a NFW profile, no Kaluza-Klein models can be tested. In
the case of a cored model, some models providing a LKP relic
density compatible with WMAP constraints can be excluded. From the
sensitivity of H.E.S.S., we derive a lower limit on the B$^{(1)}$
mass of 500 GeV.

\section{Discussion}
\label{sec:discussion}

The Sgr dSph galaxy is among the best target to search for DM
signal. Sgr dSph region is devoid of astrophysical background
unlike the Galactic Center region. Indeed, the presence of several
$\gamma$-ray emitters and diffuse emission makes difficult to
disentangle the emission from the very center from that of other
objects. The absence of gas in Sgr
offers a cleaner environment to search for $\gamma$-ray emission.\\
The DM profile uncertainties for Sgr dSph are about one order of
magnitude. In contrast, the Milky Way DM profile suffers from
large uncertainties, up to five orders of magnitude, due to the
difficulty of measuring the structural parameters of our Galaxy
in the central 10$^{-2}$ pc.\\
A modest observation time allows us for the first time to test
some pMSSM models for the $\gamma$-ray annihilation of neutralinos
in the direction of Sgr dSph. Some KK models can already be excluded.\\
Sgr dSph is a target for deeper observations by H.E.S.S. With 100
hours, H.E.S.S. will be able to test pMSSM models assuming a NFW
profile. With such an observation time, H.E.S.S. could exclude all
the cosmologically allowed KK models in case of a core profile.

\section{Conclusion}
The observations of the Sagittarius dwarf spheroidal galaxy with
H.E.S.S. performed in June 2006 reveal no significant $\gamma$-ray
excess at the nominal target position. The Sagittarius dwarf dark
matter halo has been modeled using latest measurements of its
structure parameters. Constraints have been derived on the
velocity-weighted cross section of the dark matter particle in the
framework of supersymmetric and Kaluza-Klein models.


\noindent {\bfseries Acknowledgements:} The support of the
Namibian authorities and of the University of Namibia in
facilitating the construction and operation of H.E.S.S. is
gratefully acknowledged, as is the support by the German Ministry
for Education and Research (BMBF), the Max Planck Society, the
French Ministry for Research, the CNRS-IN2P3 and the Astroparticle
Interdisciplinary Programme of the CNRS, the U.K. Particle Physics
and Astronomy Research Council (PPARC), the IPNP of the Charles
University, the South African Department of Science and Technology
and National Research Foundation, and by the University of
Namibia. We appreciate the excellent work of the technical support
staff in Berlin, Durham, Hamburg, Heildelberg, Palaiseau, Paris,
Saclay, and in Namibia in the construction and operation of the
equipment. We thank Edmond Giraud for suggestions and discussions
concerning the Sagittarius Dwarf Galaxy observations by H.E.S.S.


\newpage

\begin{table}[!htp]
\caption{Structural parameters for a cusped NFW ($\rm r_s,A$) and
a cored ($\rm r_c,v_a$) DM halo model, respectively. The values of
the solid-angle-averaged l.o.s integrated squared DM distribution
are reported in both cases for the solid angle integration region
$\rm \Delta \Omega = 2\times 10^{-5} sr$.} \label{tab:table}
\vspace{0.3cm}
\begin{center}
\begin{tabular}{|c||c|c|c|}
\hline
Halo type&Parameters&$\rm \bar{J}$&Fraction of signal \\
&&$\rm (10^{24} GeV^{2}cm^{-5})$&in $\rm\Delta\Omega = 2 \times 10^{-5}$sr\\
\hline \hline
Cusped NFW halo &r$_s$ = 0.2 kpc&$\rm 2.2$& 93.6\%\\
&$\rm A = \rm 3.3\times 10^7 M_{\odot}$&& \\
\hline
Cored halo&r$_c$ = 1.5 pc&$\rm 75.0$&99.9\%\\
&v$_a$ = 13.4 $\rm km\ s^{-1}$&&\\
\hline
\end{tabular}
\end{center}
\end{table}

\vspace{1cm}

\begin{table}[!htp]
\caption{Region of the pMSSM parameter space randomly scanned to
generate the models. A set of free parameters in the considered
ranges corresponds to a pMSSM model.} \label{tab:table2}
\vspace{0.3cm}
\begin{center}
\begin{tabular}{c}
\hline \hline
pMSSM parameter space region\\
\hline \hline
\\
100 GeV $\le \mu\le$ 30 TeV\\
100 GeV $\le M_2\le$ 50 TeV\\
50 GeV $\le M_A\le$ 10 TeV\\
100 GeV $\le m_0\le$ 1 TeV\\
$-$3 TeV $\le A_t\le$ 3 TeV\\
$-$3 TeV $\le A_b\le$ 3 TeV\\
1.2 $\le \tan \beta \le$ 60\\
\\
\hline \hline
\end{tabular}
\end{center}
\end{table}

\listoffigures
\listoftables

\end{document}